\documentclass[11pt]{article}

\usepackage[margin=1in]{geometry}
\usepackage{amsmath}
\usepackage{booktabs}
\usepackage{graphicx}
\usepackage[round]{natbib}
\usepackage[hidelinks]{hyperref}
\usepackage[margin=1in]{geometry}
\usepackage{amsmath}
\usepackage{booktabs}
\usepackage{verbatim}
\usepackage[round]{natbib}
\usepackage[hidelinks]{hyperref}
\usepackage{caption}
\newenvironment{rcode}%
  {\par\smallskip\noindent\rule{\linewidth}{0.4pt}\vspace{-0.3em}\small\verbatim}%
  {\endverbatim\vspace{-0.3em}\noindent\rule{\linewidth}{0.4pt}\par\smallskip}
 
\usepackage{caption}
\usepackage{placeins}
\graphicspath{{../R/ACTIVE/figs_OLS/}{figs_OLS/}{./}}
\newcommand{\rateunit}{percentage points}
\newcommand{\W}{\mathbf{W}}
\newcommand{\X}{\mathbf{X}}

\title{Spatial Dependence in the Self-Response:\\
       Spatial Dependence, Modeling, and Operational Consequences}
\author{Emanuel Ben-David\thanks{Any views expressed are those of
the author and not those of the U.S.\ Census Bureau.}\\ U.S.\ Census Bureau}
\date{\today}

\begin{document}
\maketitle

\begin{abstract}
\noindent
The U.S.\ Census Bureau's Low Response Score (LRS) is a central planning
instrument for identifying places likely to require additional self-response
outreach and nonresponse follow-up. The published LRS is intentionally
interpretable: it is built from tract-level covariates using an ordinary least
squares specification. That transparency, however, leaves open an important
question for official statistics: how much spatial structure remains after the
own-tract covariates have done their work, and what form does that structure
take? Using the observed 2010 Census mail non-return rate for 71,076 U.S.
census tracts and the twenty-five Erdman--Bates LRS predictors, this paper
compares the full spatial autoregressive model family under queen-contiguity
weights and validates the leading candidates with both random and spatial-block
cross-validation. OLS leaves strong residual spatial autocorrelation
($I=0.399$). Formal diagnostics and model comparisons indicate that the
remaining dependence is primarily error-type rather than a global endogenous
lag process. Although the spatial Durbin model minimizes in-sample AIC,
spatial-block validation reverses that ranking: the error-family models
(SEM/SDEM) generalize best, while the AIC-best SDM is weakest out of sample.
The SDEM provides an interpretable middle ground, absorbing residual spatial
dependence while representing neighborhood demographic effects as local
spillovers. Robustness checks show that these conclusions are invariant to the
weights definition and are not an artifact of tract-size-driven
heteroskedasticity. The results suggest that LRS-style response models should
be evaluated with spatial validation, not only in-sample fit, and that local
neighborhood context can be operationally meaningful without invoking a global
response-contagion mechanism.
\end{abstract}

{\small\raggedright
\par}
\bigskip

\section{Introduction}
\label{sec:intro}

National statistical agencies use response-propensity models to decide where
limited outreach, partnership, and field resources should be concentrated. In a
decennial census, even small improvements in identifying places likely to
self-respond at low rates can affect cost, staffing, and coverage. This is the
setting in which the U.S.\ Census Bureau developed the Low Response Score
(LRS), a tract- and block-group-level planning measure designed to locate,
predict, and manage hard-to-survey populations \citep{erdman2017lrs}. The
score's value lies partly in its transparency: unlike many high-performing
machine-learning entries in the Census return-rate prediction challenge, the
LRS is a reproducible ordinary least squares model built from a fixed set of
Planning Database predictors.

Transparency does not remove spatial dependence. Census tracts are embedded in
shared housing markets, transportation systems, media markets, local government
contexts, and histories of civic participation. Adjacent tracts can therefore
have similar response behavior even after own-tract demographic and housing
characteristics are controlled. For an operational score, this matters in two
ways. First, residual spatial autocorrelation can make conventional inference
too optimistic. Second, if the remaining spatial structure is substantively
interpretable, the score may be missing neighborhood information that outreach
programs could use. A tract may be easier or harder to count not only because of
its own characteristics, but because of the surrounding context in which those
characteristics are embedded.

This paper asks whether the residual structure left by the LRS-style OLS
specification is spatial, what spatial process best describes it, and whether
the resulting models improve prediction under validation designs that mimic
operational deployment. The distinction between spatial mechanisms is central.
A spatial lag model implies global response spillovers: a tract's outcome
depends on neighboring outcomes and changes propagate through
$(I-\rho W)^{-1}$. A spatial error model instead treats the dependence as
unobserved spatially correlated context. A spatial Durbin error model (SDEM)
adds local neighbor-covariate effects while keeping the dependence in the
error. These alternatives have very different interpretations for an official
statistical agency: the first suggests response contagion, the second omitted
regional context, and the third local neighborhood composition.

The paper makes four contributions. First, it evaluates the original LRS
covariate set against the observed mail non-return outcome rather than against
the published LRS fitted value, avoiding an algebraic exercise in reproducing
the original OLS score. Second, it applies the nested spatial autoregressive
model family to a national tract-level response outcome and uses formal
diagnostics to distinguish lag, error, and Durbin mechanisms. Third, it shows
that spatial-block validation can overturn the in-sample information-criterion
ranking: the AIC-best spatial Durbin model does not generalize best to
withheld geography. Fourth, it separates point-estimate robustness from
inference under tract-size-driven heteroskedasticity, a practical concern for
rates built from heterogeneous tract populations and ACS-derived predictors.

The analysis is anchored to the 2010 mail non-return outcome because that is
the outcome on which the published Erdman--Bates LRS was defined and the only
setting in which the twenty-five original predictors, their transformations,
and the official score can be compared without changing the estimand. The
contemporary Planning Database now includes a 2020-based Low Response Score
that reflects a broader self-response environment, including internet response
and contact-strategy changes \citep{census2026pdbupdate}. That evolution makes
the present exercise more, not less, important: before extending the score to
new modes and vintages, it is useful to know what kind of spatial dependence is
left by the interpretable tract-level model in the original LRS setting. The
conclusion returns to how the same validation design should be applied to the
2020-based score.

The remainder of the paper is organized as follows.
Section~\ref{sec:data-methods} describes the data, outcome, spatial weights,
model family, and validation design. Section~\ref{sec:model-comparison}
reports the specification search, impact decomposition, validation results, and
robustness checks. Section~\ref{sec:discussion} discusses the contribution,
limitations, and implications for future LRS revisions.

\section{Data, Models, and Validation Design}
\label{sec:data-methods}

The Low Response Score \citep{erdman2017lrs} is estimated by ordinary least
squares (OLS) on twenty-five tract-level predictors drawn from the Planning
Database. Because the predictors and the published score are deterministic
functions of one another, this paper models the \emph{observed} 2010 mail
non-return rate directly, on the logit scale, so that a residual exists that can
carry spatial structure. Throughout,
$y_i = \operatorname{logit}(\text{non-return}_i)$ for the $n = 71{,}076$ tracts
of the United States (the fifty states and the District of Columbia, including
Alaska and Hawaii), $X$ collects the Erdman--Bates predictors with their
published transformations, and $W$ is the row-standardized queen-contiguity
weights matrix.

The analytic sample contains the tracts for which the observed mail non-return
rate and all twenty-five LRS predictors are available. Tracts with zero or
suppressed denominators, missing ACS inputs, or incompatible geography are
excluded. The predictors retain the Erdman--Bates transformations so that the
spatial specifications isolate the contribution of spatial structure rather
than a redesigned covariate set.

We compare OLS with the standard spatial autoregressive lattice: the spatial
lag of $X$ (SLX), the spatial lag model (SLM), the spatial error model (SEM),
the spatial Durbin model (SDM), the spatial Durbin error model (SDEM), and the
combined lag-and-error model (SARAR). The general form is
\[
y = \rho Wy + X\beta + WX\theta + u,
\qquad
u = \lambda Wu + \varepsilon ,
\]
with restrictions on $\rho$, $\theta$, and $\lambda$ defining each named model.
The primary weights matrix uses row-standardized queen contiguity, so $Wy$ and
$WX$ are neighbor averages. A $k=8$ nearest-neighbor matrix provides a robustness
check that removes variation in graph degree.

Model choice is evaluated in three stages. First, Moran's $I$ and robust
Rao-score diagnostics are applied to the OLS residuals to identify whether the
remaining dependence is lag- or error-type. Second, likelihood, information
criteria, residual Moran's $I$, and restriction tests compare the fitted SAR
family in sample. Third, the leading models are evaluated out of sample under
both random folds and spatial-block folds. Random folds measure local
interpolation when held-out tracts are surrounded by training geography.
Spatial-block folds withhold contiguous regions and are therefore a more
conservative proxy for deployment to genuinely unsampled geography
\citep{roberts2017cross}. The gap between random and spatial-block performance
is interpreted as a measure of neighbor reliance.

This design tests four claims: (i) the unmodeled structure left by OLS is
spatial and error-type; (ii) within the spatial autoregressive family, the SEM
and SDEM are preferred on formal and substantive grounds; (iii) out-of-sample
validation overturns the in-sample information-criterion ranking, exposing the
SDM as an in-sample artifact; and (iv) these conclusions are invariant to the
weights definition and are not driven by residual heteroskedasticity.

\section{Results: Specification Search and Model Comparison}
\label{sec:model-comparison}

\subsection{The geography of mail non-response}
\label{sec:geography}

\begin{figure}[!htbp]
\centering
\includegraphics[width=0.92\textwidth]{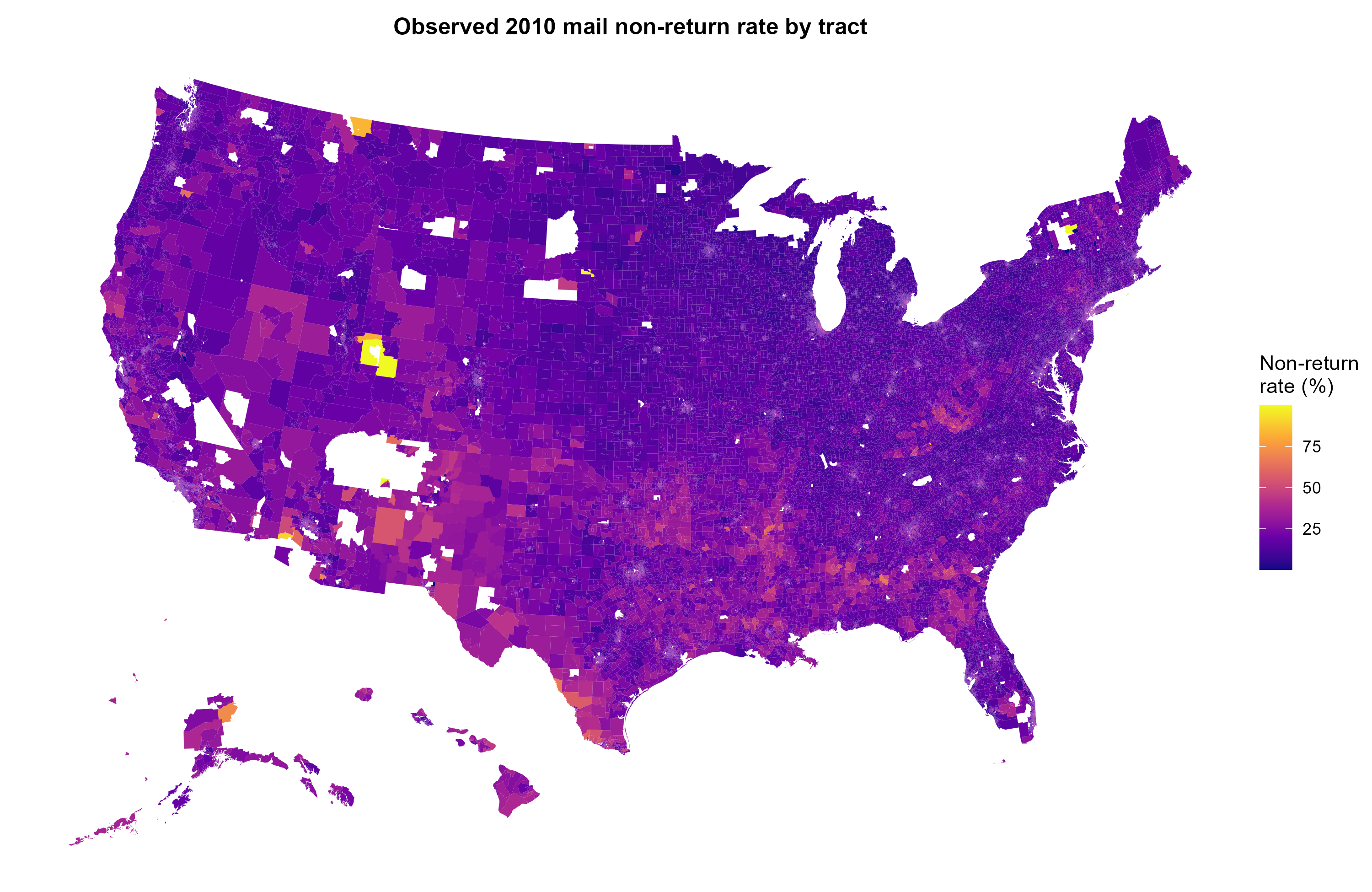}
\caption{Observed 2010 mail non-return rate by census tract (the modeling
outcome, shown on the rate scale). Darker tracts return their forms at high
rates; brighter tracts are harder to count. The rate clusters into coherent
regions---the rural South, the southern border, the desert Southwest, and urban
cores---rather than varying tract by tract. This spatial clustering, left
unmodeled by an own-tract OLS specification, is the structure the spatial models
are designed to capture. (Alaska and Hawaii are repositioned for
display.)}
\label{fig:observed}
\end{figure}

Figure~\ref{fig:observed} maps the outcome. Color encodes the observed tract
non-return rate, from dark (low non-return, easy to count) to bright yellow
(high non-return, hard to count). Two features motivate everything that
follows. First, the rate is not noise sprinkled across the map: it forms
coherent regional patches---elevated across the rural South, the
Texas--Mexico border, parts of the desert Southwest and tribal lands, and in
dense urban cores---so that neighboring tracts tend to share levels. Second,
most of the country sits in a narrow low range (roughly $15$--$30\%$),
punctuated by these high-non-return clusters. Spatial clustering of this kind is
precisely what an OLS model built on own-tract covariates cannot represent: if
two adjacent tracts with identical covariates nonetheless differ because they
sit in different regional contexts, that difference is pushed into the residual.
The remainder of the analysis asks whether modeling that spatial residual
improves the score, and if so, through which mechanism.

\FloatBarrier

\subsection{Spatial diagnostics and specification search}
\label{sec:spec-search}

The OLS residuals are strongly spatially clustered: Moran's $I = 0.399$
($p < 10^{-16}$). To identify \emph{which} spatial process generates this
clustering we apply the Anselin--Florax robust Rao-score (Lagrange multiplier)
tests \citep{anselin1996simple}, which discriminate a spatial \emph{lag}
process from a spatial \emph{error} process at the OLS point. The robust error
statistic dominates the robust lag statistic by a factor of roughly
twenty-six under queen contiguity ($9{,}319$ vs.\ $357$, both
$p < 10^{-16}$); the decision rule nominates an error-type process.

We then estimate the full nesting lattice---OLS, the spatial lag of $X$ (SLX),
the spatial lag model (SLM), the spatial error model (SEM), the spatial Durbin
model (SDM), the spatial Durbin error model (SDEM), and the combined
lag-and-error model (SARAR)---and test the restrictions among them
(Table~\ref{tab:spec}). All likelihood-ratio (LR) tests reject at
$p < 10^{-16}$: SEM improves on OLS ($\mathrm{LR}=22{,}544$, $\mathrm{df}=1$),
the Durbin terms add information beyond the error structure
(SDEM vs.\ SEM: $\mathrm{LR}=1{,}102$, $\mathrm{df}=25$) and beyond the lag
structure (SDM vs.\ SLM: $\mathrm{LR}=6{,}084$, $\mathrm{df}=25$), and the
error term adds information beyond lagged covariates alone
(SDEM vs.\ SLX: $\mathrm{LR}=20{,}916$, $\mathrm{df}=1$). Burridge's
common-factor test \citep{burridge1981testing} rejects the restriction that the
spatial Durbin model collapses to a pure error model ($\mathrm{LR}=1{,}409$,
$\mathrm{df}=25$), so the spatially lagged covariates carry information that a
parsimonious error process does not; and the spatial Hausman test
\citep{pace2008spatial} rejects equality of the OLS and SEM coefficient vectors
($\chi^2 = 1{,}421$, $\mathrm{df}=26$), the formal signal that a pure error
model leaves systematic structure---here, the neighbor covariates---unmodeled.

\begin{table}[!htbp]
\centering
\caption{Specification comparison on $y=\operatorname{logit}(\text{non-return})$,
queen contiguity, $n=71{,}076$. $\Delta$AIC is relative to the
information-criterion-best model (SDM). Residual Moran's $I$ near zero indicates
spatial structure absorbed.}
\label{tab:spec}
\begin{tabular}{lrrrrrr}
\toprule
Model & $\log L$ & AIC & $\Delta$AIC & $\rho$ & $\lambda$ & Resid.\ $I$ \\
\midrule
OLS   & -18505 & 37063 & 23901 & --     & --     &  0.399 \\
SLX   & -17140 & 34383 & 21221 & --     & --     &  0.392 \\
SLM   &  -9570 & 19197 &  6034 &  0.505 & --     &  0.054 \\
SEM   &  -7233 & 14522 &  1359 & --     &  0.651 & -0.045 \\
SDM   &  -6528 & 13162 &     0 &  0.623 & --     & -0.041 \\
SDEM  &  -6682 & 13469 &   307 & --     &  0.626 & -0.040 \\
SARAR &  -6916 & 13890 &   728 & -0.312 &  0.808 & -0.043 \\
\bottomrule
\end{tabular}
\end{table}
\FloatBarrier

\begin{figure}[!htbp]
\centering
\includegraphics[width=0.72\textwidth]{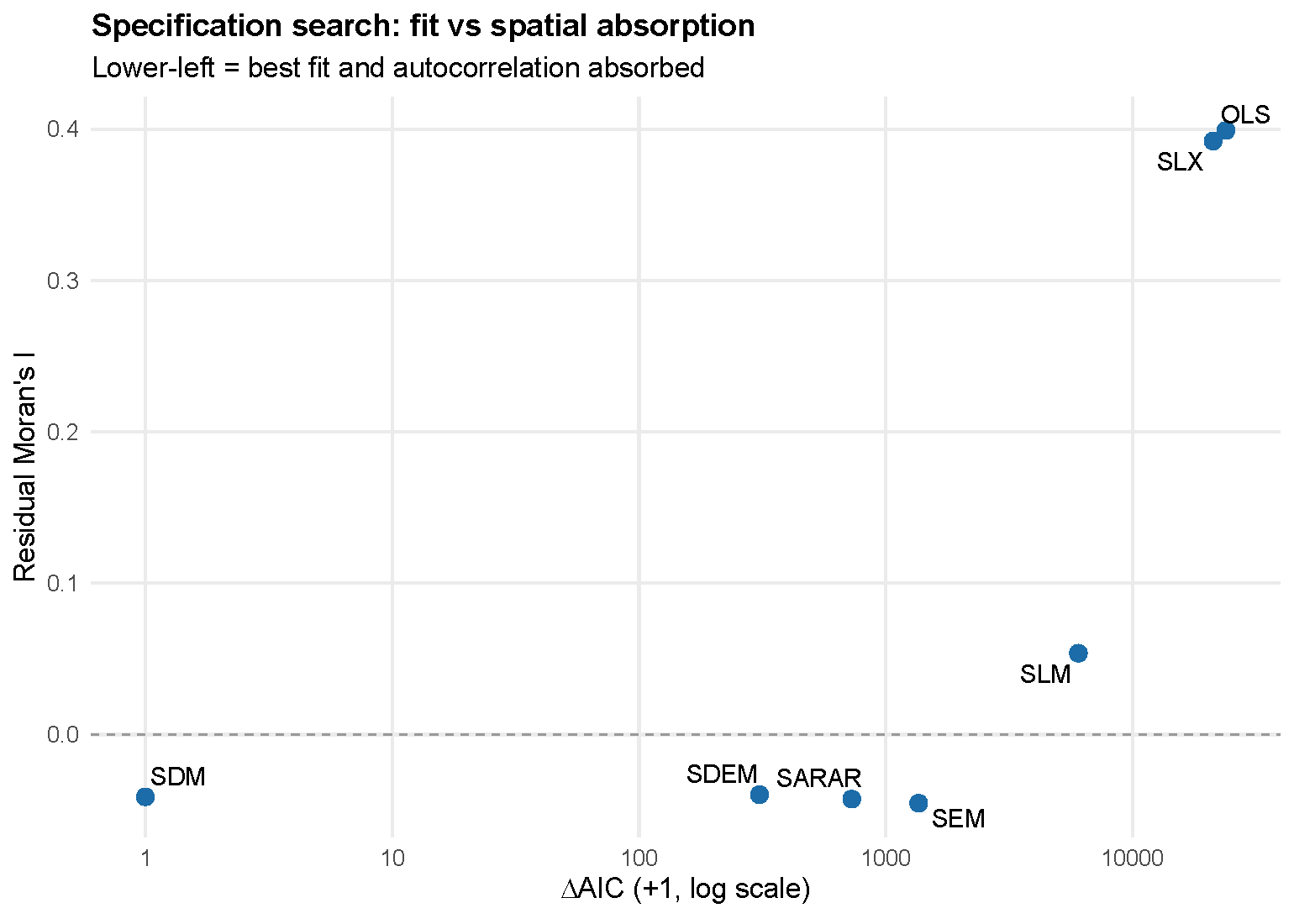}
\caption{The specification search in one plane: in-sample fit ($\Delta$AIC, log
scale; lower is better) versus residual spatial autocorrelation (Moran's $I$;
zero is fully absorbed). The desirable region is the lower-left. OLS and SLX
fit poorly and leave large residual autocorrelation; the pure lag model (SLM)
fits better but still leaves $I=0.05$; the error and Durbin models (SEM, SDM,
SDEM, SARAR) cluster in the lower-left, fitting best and absorbing the spatial
structure. The contrast between SLX and the autoregressive models shows that
lagged covariates improve fit but only a $\rho$ or $\lambda$ term removes the
autocorrelation.}
\label{fig:spec}
\end{figure}

Figure~\ref{fig:spec} renders the search in a single plane: in-sample fit on the
horizontal axis ($\Delta$AIC, plotted on a log scale so that the very wide range
is legible; lower is better) against residual spatial autocorrelation on the
vertical axis (Moran's $I$; zero means the spatial structure has been fully
absorbed). A good model sits in the lower-left corner---low $\Delta$AIC and
near-zero residual autocorrelation. OLS and SLX occupy the upper-right: poor fit
and a large amount of leftover spatial structure ($I \approx 0.4$). Adding a
pure lag term (SLM) improves fit substantially but still leaves positive
residual autocorrelation ($I = 0.054$). Only the models containing a $\rho$ or
$\lambda$ autoregressive term---SEM, SDM, SDEM, and SARAR---reach the
lower-left, simultaneously fitting best and driving residual autocorrelation
slightly below zero. The figure makes two points at once: the lagged covariates
in SLX improve fit but barely touch the autocorrelation, whereas the
autoregressive term is what collapses it; and the error and Durbin models form a
tight cluster of well-fitting, fully-absorbing specifications from which the
final choice must be made.

\FloatBarrier

Two further features of Table~\ref{tab:spec} shape the analysis. SARAR returns a
negative autoregressive parameter ($\rho = -0.312$) alongside a large error
parameter ($\lambda = 0.808$): the textbook signature of a spurious lag term in
an error-driven process \citep{elhorst2010applied}, so we retain SARAR as a
diagnostic only. And the information-criterion ranking favors SDM, a result
Section~\ref{sec:validation} shows does not survive out-of-sample evaluation.

\subsection{Spatial spillovers}
\label{sec:impacts}

\begin{figure}[!htbp]
\centering
\includegraphics[width=\textwidth]{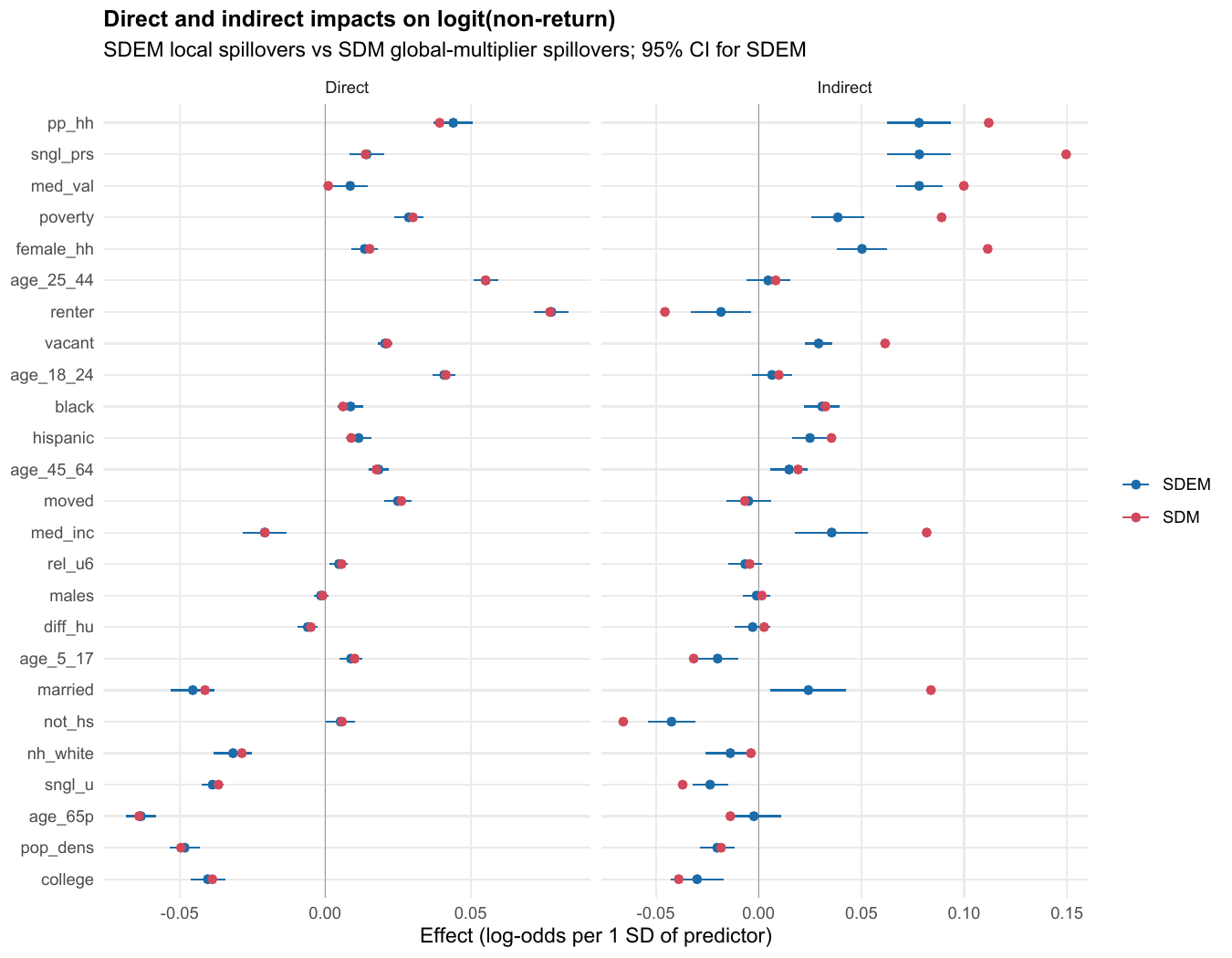}
\caption{Direct (own-tract) and indirect (neighbor) impacts of each predictor on
logit(non-return), per one standard deviation, for SDEM (blue, with 95\%
intervals) and SDM (red). Predictors are ordered by SDEM total effect. The
direct effects coincide across the two models; the indirect effects are
systematically larger under SDM because its endogenous lag propagates each
covariate through the global multiplier $(I-\rho W)^{-1}$, whereas an SDEM
indirect effect is the local lagged coefficient $\theta$ with no feedback. For
several predictors---median home value (med\_val), single-person households
(sngl\_prs), persons per household (pp\_hh), female-headed households
(female\_hh)---the neighbor effect rivals or exceeds the own-tract effect.}
\label{fig:impacts}
\end{figure}

Because the Durbin specifications include spatially lagged covariates, their
coefficients do not have a direct marginal interpretation; we report the
direct, indirect, and total impacts \citep{lesage2009introduction}.
Figure~\ref{fig:impacts} contrasts the SDM and SDEM decompositions, split into
the own-tract (direct) effect and the neighbor (indirect) effect of each
predictor. Reading the figure: each row is a predictor (ordered by SDEM total
effect, largest at the top), the horizontal position is the estimated effect in
log-odds per one standard deviation of the predictor, blue markers are SDEM
with 95\% intervals, and red markers are SDM. In the left panel (direct
effects) the blue and red markers sit essentially on top of one another: the
own-tract effects are the same regardless of which spatial mechanism is assumed,
so the substantive own-tract story is robust to the modeling choice. In the
right panel (indirect effects) the red SDM markers lie systematically farther
from zero than the blue SDEM markers. This is the global multiplier:
SDM propagates each covariate's influence through $(I-\rho W)^{-1}$, so its
spillovers accumulate across the whole map, whereas an SDEM spillover is simply
the lagged coefficient $\theta$---a local, first-order, neighbor-only effect.
Because both models leave the same residual autocorrelation
(Table~\ref{tab:spec}), the larger SDM spillovers reflect the assumed mechanism
rather than additional structure absorbed, and the local SDEM spillovers are the
more defensible estimate for an official-statistics application.

\FloatBarrier

The decomposition yields a substantive finding obscured by a single OLS
coefficient: for several predictors the neighborhood effect rivals or exceeds
the own-tract effect. Median home value has a negligible own-tract effect but a
large, highly significant neighborhood effect; single-person households,
female-headed households, and the non-Hispanic Black share show the same
pattern. For median household income and married-couple households the direct
and indirect effects carry opposite signs---higher own-tract income lowers
non-response while higher surrounding income raises it---so the total effect
nearly cancels. For mail non-response, the composition of a tract's neighbors
can matter as much as the tract's own composition.

\begin{figure}[!htbp]
\centering
\includegraphics[width=0.92\textwidth]{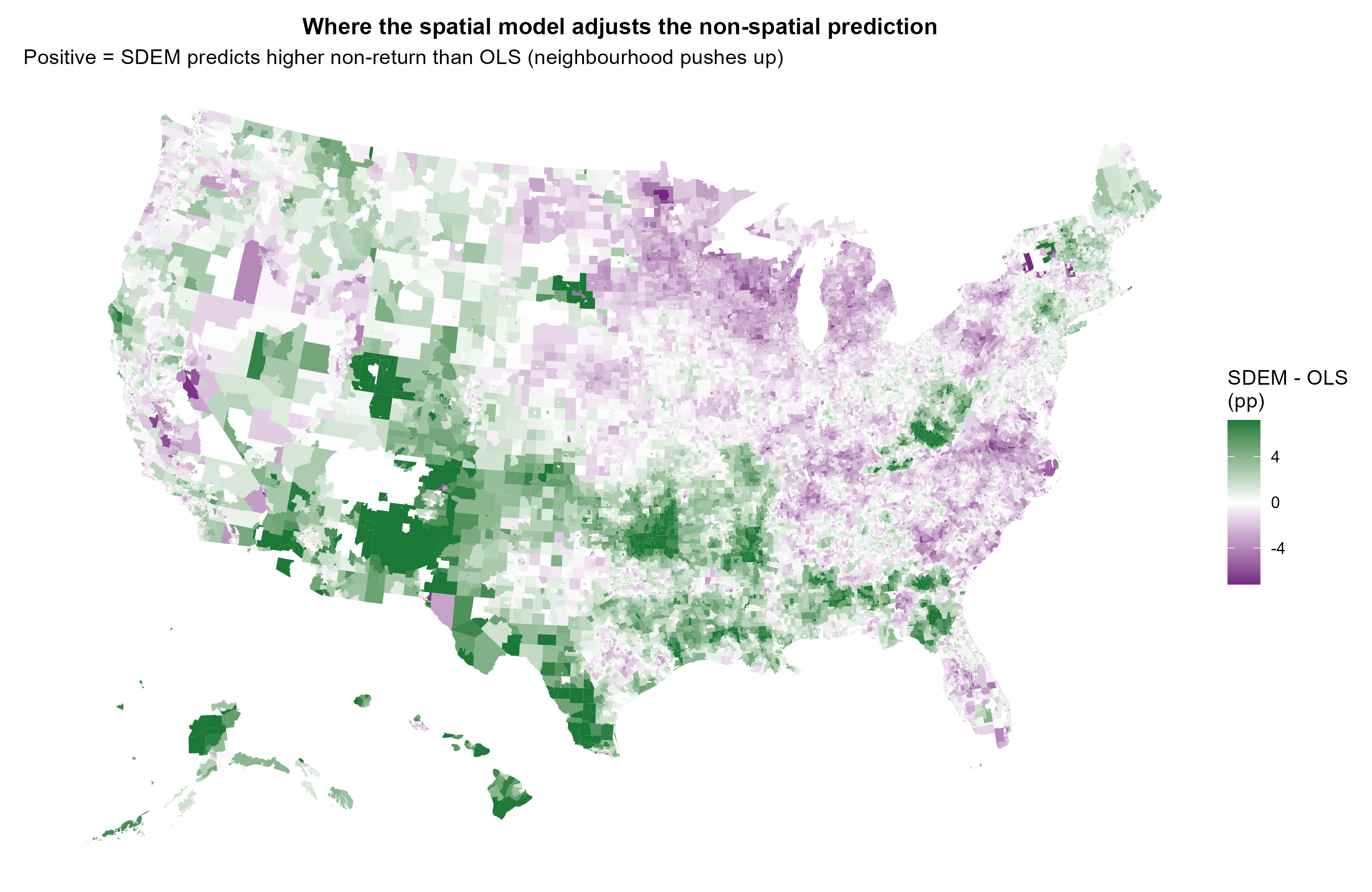}
\caption{The spatial model's adjustment to the non-spatial prediction
(SDEM minus OLS predicted non-return rate, \rateunit). Green: SDEM predicts a
higher non-return rate than OLS because the neighborhood is harder to count than
own-tract covariates imply; purple: the reverse. The adjustments form coherent
regions---raised across Texas and the south-central states, lowered across the
upper Midwest and Northeast---rather than tract-level noise, visualizing the
regional information the spatial model adds.}
\label{fig:adjust}
\end{figure}

These spillovers are not an abstraction; they have a geography.
Figure~\ref{fig:adjust} maps the difference between the SDEM and OLS predicted
non-return rates, that is, how much the spatial model moves the own-tract
prediction once neighborhood information is admitted. Green tracts are those
where SDEM predicts a \emph{higher} non-return rate than OLS---the surrounding
context is harder to count than the tract's own covariates imply---and purple
tracts are the reverse. The adjustments are not random tract-level jitter: they
form broad, coherent regions. SDEM raises predictions across much of Texas, the
south-central states, and parts of the Southwest and Central Valley, and lowers
them across the upper Midwest and Great Lakes, Appalachia, and stretches of the
Northeast. This is exactly the regional signal that an own-tract model cannot
see, made visible: where the map is strongly colored, a tract's neighbors carry
information about its response propensity that its own characteristics miss.

\FloatBarrier

\subsection{Spatial fit and a District of Columbia case study}
\label{sec:fit}

\begin{figure}[!htbp]
\centering
\includegraphics[width=\textwidth]{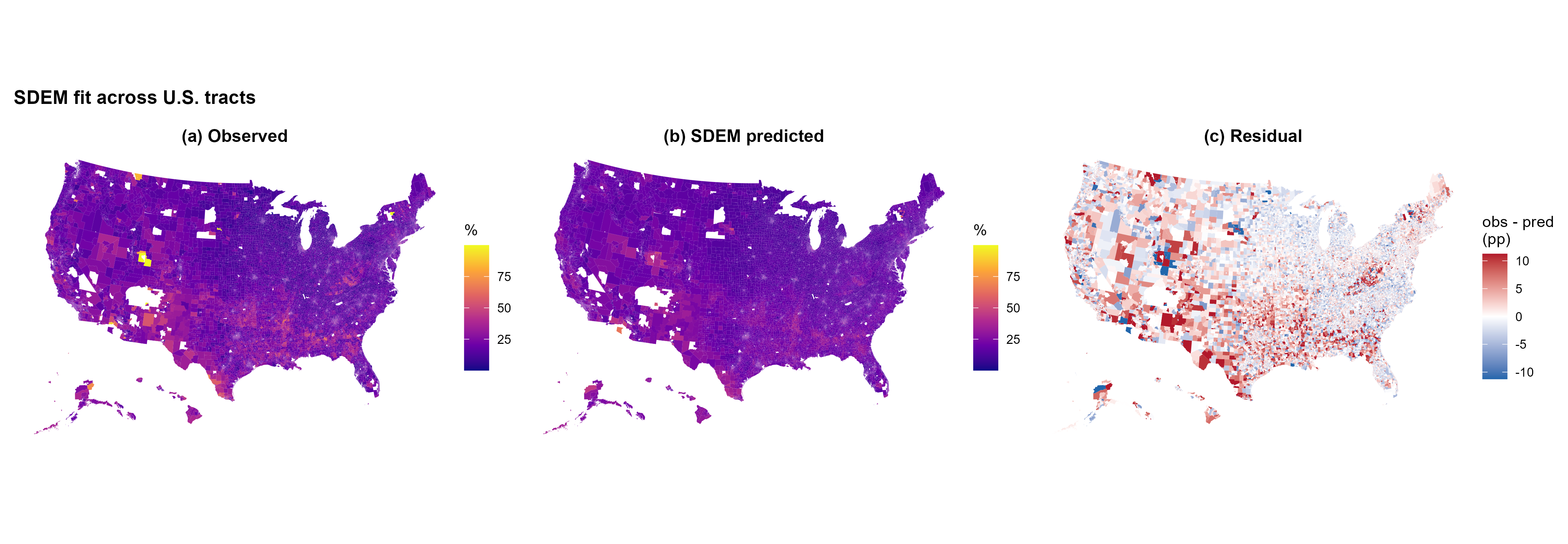}
\caption{SDEM fit across U.S.\ tracts: (a) observed non-return rate, (b) SDEM
predicted rate (same color scale), and (c) residual (observed minus predicted,
\rateunit; red = under-prediction, blue = over-prediction). Predicted closely
reproduces observed. The residual map shows small, spatially unstructured
errors---the salt-and-pepper texture corresponding to the near-zero residual
Moran's $I$---confirming that the spatial structure has been absorbed.}
\label{fig:usfit}
\end{figure}

Figure~\ref{fig:usfit} shows the model's in-sample fit across all tracts in three panels:
the observed rate (a), the SDEM predicted rate (b), and the residual (c, observed
minus predicted, in \rateunit). Panels (a) and (b) are nearly
indistinguishable, which is the point: the model reproduces the broad national
geography of non-response, including the elevated South and border regions. The
residual panel (c) is dominated by small, alternating positive and negative
values without a strong large-scale pattern---a salt-and-pepper texture rather
than coherent regional blocks---confirming visually what the residual Moran's
$I$ of $-0.04$ states numerically: the spatial structure has been absorbed and
what remains is close to spatially unstructured local variation. What little
residual coherence remains is faint and local, with somewhat larger errors in
the sparsely populated West that are examined in
Section~\ref{sec:robustness}.

\FloatBarrier

\begin{figure}[!htbp]
\centering
\includegraphics[width=\textwidth]{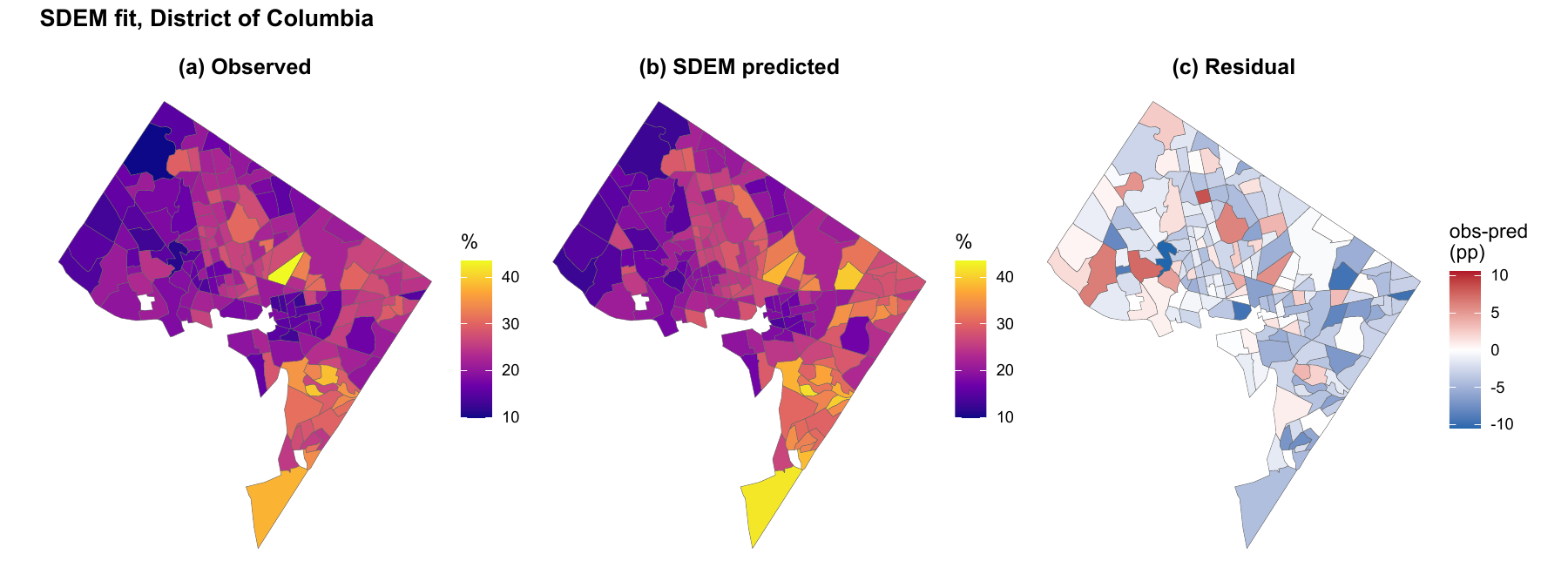}
\caption{SDEM fit for the District of Columbia: (a) observed, (b) predicted,
(c) residual (\rateunit). DC's east--west non-response gradient---low in the
northwest, high east of the Anacostia---is reproduced by the model; residuals
are modest and show no remaining systematic gradient.}
\label{fig:dcfit}
\end{figure}

The District of Columbia, a single dense and well-studied jurisdiction, makes
the same points at a scale where individual tracts are visible.
Figure~\ref{fig:dcfit} repeats the observed/predicted/residual triptych for DC.
The well-known east--west gradient is clear in the observed panel: low
non-return in the affluent, high-response northwest and markedly higher
non-return in the historically harder-to-count wards east of the Anacostia
River. The SDEM prediction reproduces this gradient, and the residuals are
modest and of both signs, with no systematic east--west pattern remaining.

\FloatBarrier

\begin{figure}[!htbp]
\centering
\includegraphics[width=0.62\textwidth]{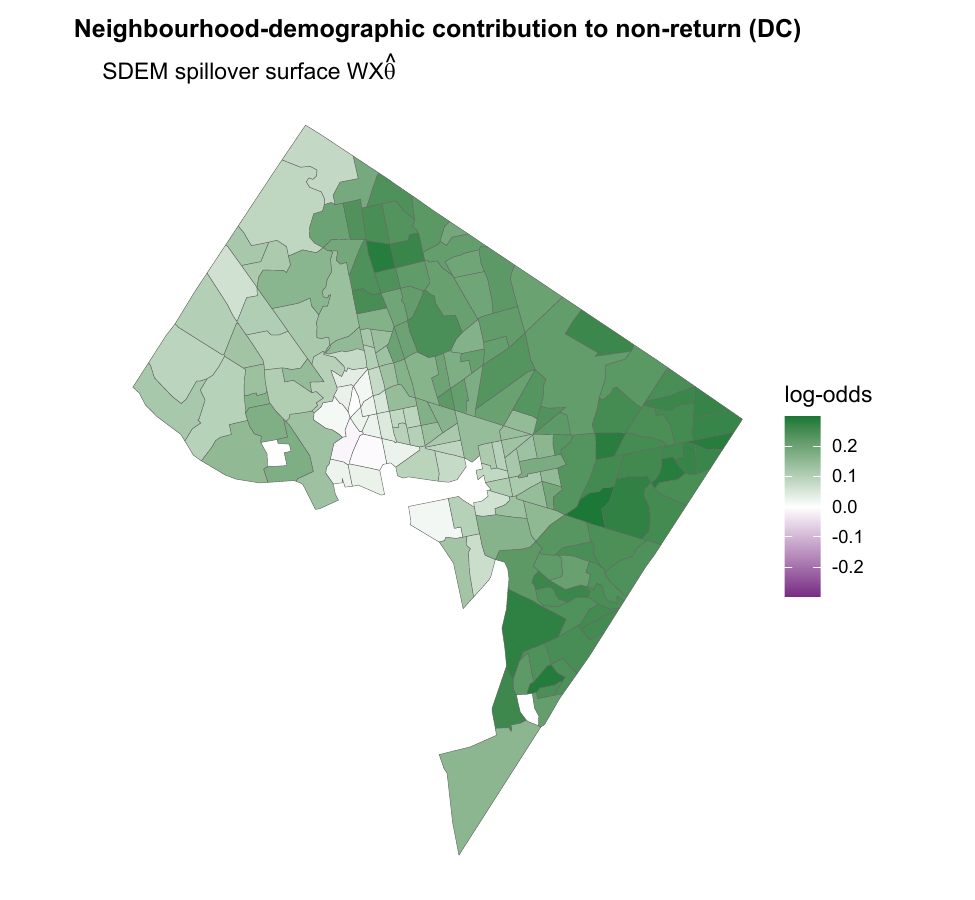}
\caption{The SDEM neighbor-demographic contribution $WX\hat\theta$ (log-odds) for
the District of Columbia: the portion of each tract's predicted non-return that
derives from its neighbors' characteristics. The contribution is positive and
largest across the eastern and southeastern wards and near zero in the central
northwest, showing that the spillover is spatially structured and reinforces the
city's east--west non-response gradient.}
\label{fig:dcneigh}
\end{figure}

Finally, Figure~\ref{fig:dcneigh} isolates the SDEM spillover surface for DC:
the neighbor-demographic contribution $WX\hat\theta$ to each tract's
non-return log-odds, that is, the part of the prediction that comes from the
\emph{neighbors'} characteristics rather than the tract's own. Green indicates a
positive neighbor contribution (surrounding demographics raise the predicted
non-return), and the contribution is itself spatially patterned: it is largest
across the eastern and southeastern wards and near zero in the central
northwest. The neighborhood spillover is therefore not a uniform city-wide
offset but a structured surface that reinforces the same east--west gradient
seen in the outcome---concrete, tract-level evidence that for these communities
the surrounding context contributes measurably to response propensity.

\FloatBarrier

\subsection{Out-of-sample validation}
\label{sec:validation}

\begin{figure}[!htbp]
\centering
\includegraphics[width=0.92\textwidth]{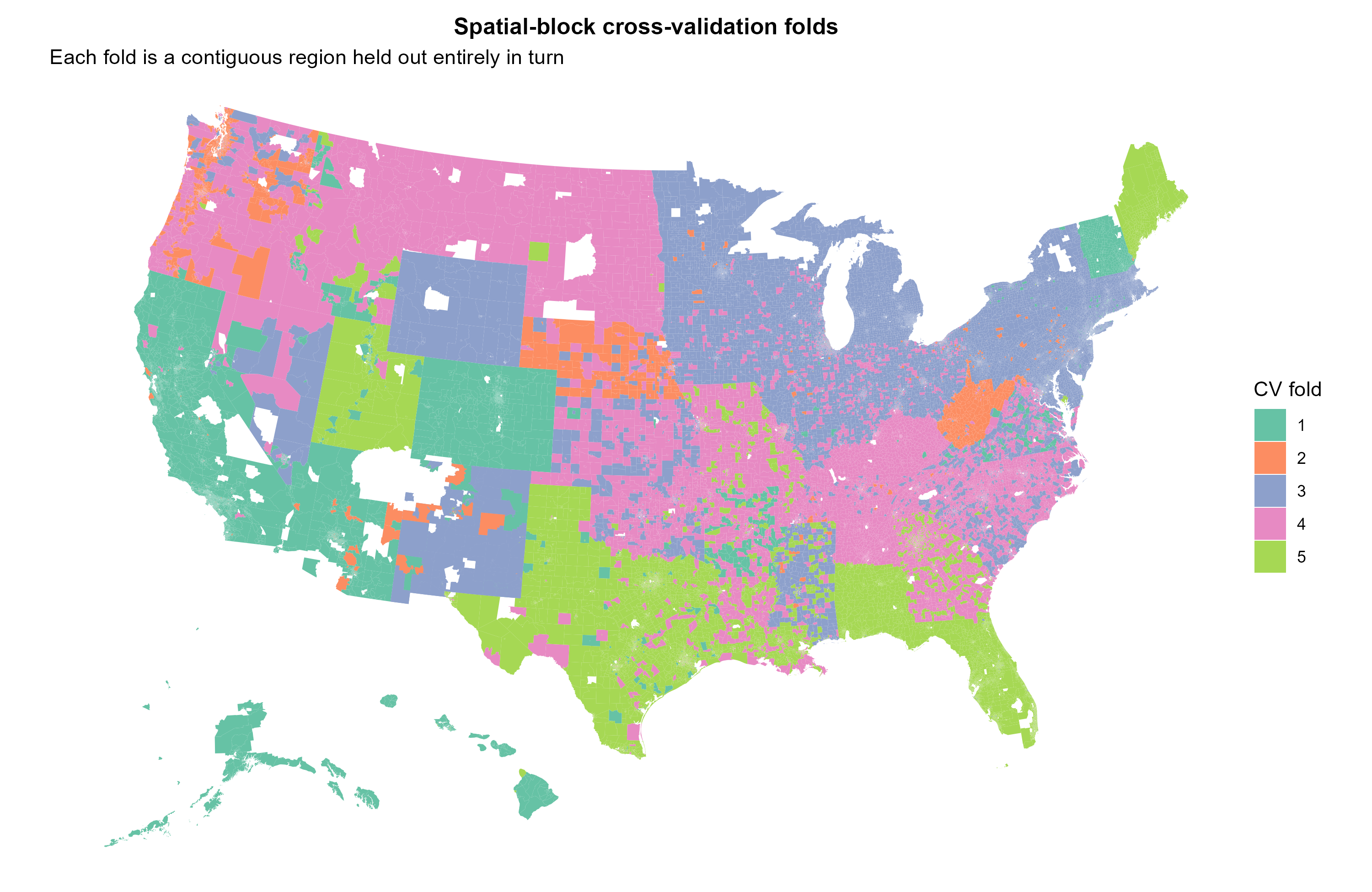}
\caption{Spatial-block cross-validation folds: each color is one of five
contiguous regions, withheld in its entirety in turn. Because a held-out tract's
neighbors are withheld with it, no local information leaks from training to test
---a conservative estimate of skill on unsampled geography, in contrast to
random folds, where a held-out tract is surrounded by training tracts.}
\label{fig:folds}
\end{figure}

Information criteria reward in-sample fit and, with the many parameters of the
Durbin specifications, can favor a model that does not generalize. Because the
score is an operational instrument applied to tracts whose response is not yet
observed, we evaluate predictive skill out of sample under two cross-validation
regimes. \emph{Spatial-block} folds hold out entire contiguous regions and
estimate skill on genuinely unsampled geography \citep{roberts2017cross};
\emph{random} folds hold out interspersed tracts and therefore leak through
autocorrelation. Figure~\ref{fig:folds} shows the spatial-block partition: each
color is one fold, a single contiguous region withheld in its turn, so that a
held-out tract's neighbors are also held out and cannot leak training
information. This is a far more demanding test than random holdout, in which a
withheld tract is typically surrounded by training tracts. The gap between the
two regimes measures how much a model leans on local neighbors versus
transferable covariate structure.

\FloatBarrier

\begin{figure}[!htbp]
\centering
\includegraphics[width=0.80\textwidth]{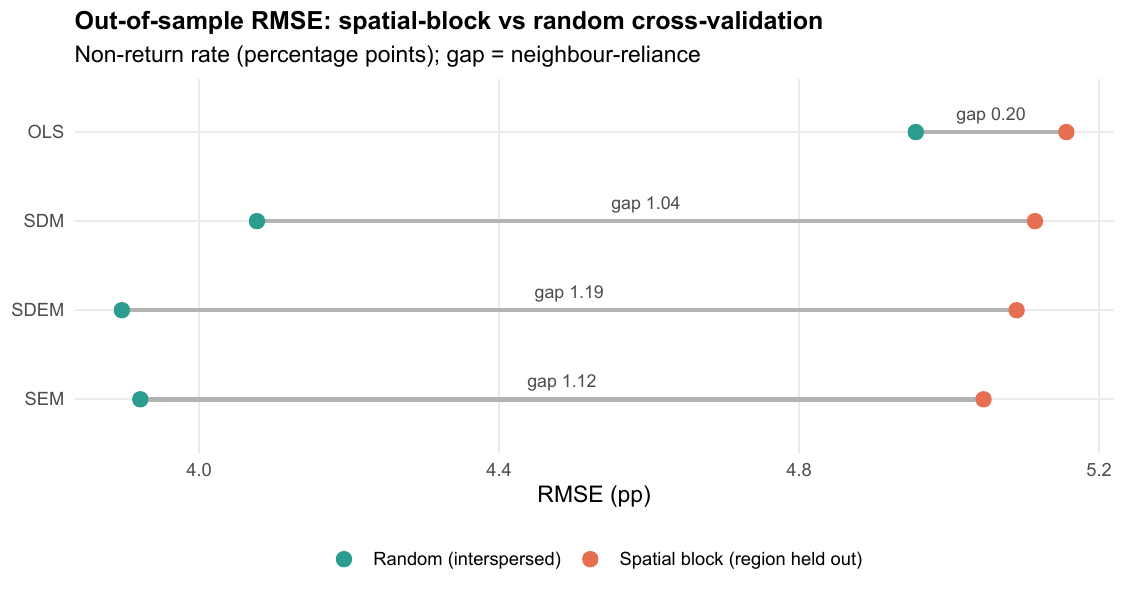}
\caption{Out-of-sample RMSE (\rateunit) for each model under random
(interspersed) and spatial-block (region held out) cross-validation; the line
length is the neighbor-reliance gap. OLS has a negligible gap (no spatial term
to lose). The spatial models predict much better under random folds than under
spatial blocks, a gap near $1$~pp, indicating that much of their apparent skill
is local interpolation rather than transferable structure. Under the demanding
spatial-block regime the error models (SEM, SDEM) lead and the AIC-best SDM
trails.}
\label{fig:cv}
\end{figure}

Held-out tracts are predicted from their training neighbors only, using the
optimal SAR predictors of \citet{goulard2017predictions}: a reduced-form block
solve for the lag family and the Gaussian Markov random field conditional mean
for the error family. Table~\ref{tab:cv} reports the back-transformed
non-return rate error in \rateunit, and Figure~\ref{fig:cv} displays it: for
each model, a marker for the random-fold RMSE and a marker for the spatial-block
RMSE, joined by a line whose length is the neighbor-reliance gap.

\begin{table}[!htbp]
\centering
\caption{Five-fold cross-validated prediction error for the non-return rate
(\rateunit). ``Gap'' is spatial $-$ random RMSE, a measure of neighbor-reliance.
Logit-scale metrics give the identical ordering.}
\label{tab:cv}
\begin{tabular}{lrrrrr}
\toprule
& \multicolumn{3}{c}{RMSE} & \multicolumn{2}{c}{MAE} \\
\cmidrule(lr){2-4}\cmidrule(lr){5-6}
Model & Spatial & Random & Gap & Spatial & Random \\
\midrule
SEM   & 5.046 & 3.922 & 1.124 & 3.666 & 2.779 \\
SDEM  & 5.090 & 3.897 & 1.193 & 3.703 & 2.763 \\
SDM   & 5.115 & 4.077 & 1.037 & 3.728 & 2.929 \\
OLS   & 5.156 & 4.956 & 0.201 & 3.757 & 3.600 \\
\bottomrule
\end{tabular}
\end{table}
\FloatBarrier

\FloatBarrier

The out-of-sample ranking inverts the information-criterion ranking. SDM, which
minimized AIC, is the \emph{worst} of the three spatial models on spatial-block
RMSE; its in-sample advantage reflects parameters that do not transfer to new
geography. The error family outperforms the lag family under both regimes,
consistent with the robust-LM verdict of Section~\ref{sec:spec-search}. SEM and
SDEM are statistically indistinguishable on spatial-block skill (a $0.04$~pp
difference, well inside the fold-to-fold variation, which ranges from $4.2$ to
$5.7$~pp): the Durbin terms aid local interpolation (SDEM narrowly leads under
random folds) but confer no advantage when an entire region is withheld. As
Figure~\ref{fig:cv} makes plain, every spatial model loses roughly a full
percentage point of RMSE moving from random to spatial holdout, while OLS loses
almost nothing---the spatial models buy their skill partly by borrowing from
neighbors, and that loan is called in when a whole region is unsampled. SDEM's
gap is the largest ($1.19$), the mirror image of its random-fold advantage:
adding lagged covariates makes the model lean hardest on local structure.

The spatial parameters are estimated consistently across folds (standard
deviations of $\lambda$ and $\rho$ below $0.013$ in every regime), so no model
rides a single fold's idiosyncrasy. Their means rise markedly under spatial
blocking relative to random folds (e.g., SDEM $\lambda = 0.619$ vs.\ $0.558$;
SEM $\lambda = 0.645$ vs.\ $0.579$): when whole regions are withheld, the model
leans harder on the spatial term to compensate for covariate structure it can no
longer borrow locally.

\subsection{Robustness}
\label{sec:robustness}

\paragraph{Weights definition.} Re-estimating the lattice under
$k=8$ nearest-neighbor weights leaves every structural conclusion intact
(Table~\ref{tab:wrobust}): the model ordering by AIC is identical, the
robust-LM verdict still favors the error process (now by a factor of
sixty-seven), residual Moran's $I$ is still driven to approximately $-0.03$ by
the error and Durbin models, and SARAR remains pathological
($\rho = -0.353$). The only systematic difference is that all spatial
parameters are higher under $k$-nearest-neighbor weights---an expected
consequence of that graph's fixed, symmetric, and generally higher degree---with
signs, relative magnitudes, and significance unchanged.

\begin{table}[!htbp]
\centering
\caption{Weights robustness: queen contiguity vs.\ $k=8$ nearest neighbors. The
spatial parameter is $\lambda$ for the error models (SEM, SDEM) and $\rho$ for
the lag model (SDM). Ordering and residual absorption are invariant; parameters
inflate uniformly under $k$NN.}
\label{tab:wrobust}
\begin{tabular}{lrrrrrr}
\toprule
& \multicolumn{2}{c}{AIC} & \multicolumn{2}{c}{Spatial param.} & \multicolumn{2}{c}{Resid.\ $I$} \\
\cmidrule(lr){2-3}\cmidrule(lr){4-5}\cmidrule(lr){6-7}
Model & Queen & $k$NN & Queen & $k$NN & Queen & $k$NN \\
\midrule
SEM  & 14522 & 14791 & 0.651 & 0.712 & -0.045 & -0.032 \\
SDM  & 13162 & 13497 & 0.623 & 0.674 & -0.041 & -0.028 \\
SDEM & 13469 & 13721 & 0.626 & 0.678 & -0.040 & -0.027 \\
\bottomrule
\end{tabular}
\end{table}
\FloatBarrier

\begin{figure}[!htbp]
\centering
\includegraphics[width=0.72\textwidth]{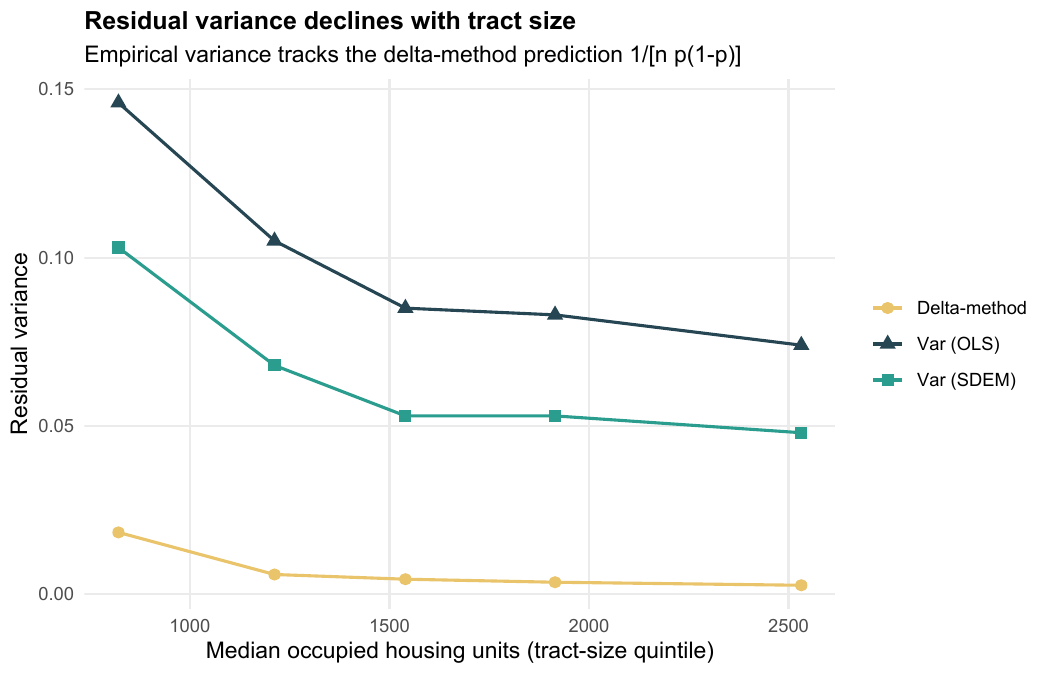}
\caption{Residual variance by tract-size quintile (median occupied housing
units), for OLS and SDEM residuals, with the delta-method prediction
$1/[n\,p(1-p)]$. Variance falls about twofold from the smallest to the largest
tracts, tracking the delta-method curve. SDEM lies below OLS (lower overall
variance) but parallel to it (the same size gradient), showing that the spatial
error model does not remove the heteroskedasticity.}
\label{fig:hetero}
\end{figure}

\paragraph{Heteroskedasticity.} The SAR family is estimated by Gaussian
maximum likelihood under a homoskedastic innovation; the error models whiten
spatial correlation but do not address heteroskedasticity. Two channels make the
innovation heteroskedastic here. The outcome, a logit-transformed rate, has
delta-method variance proportional to $1/[\,n_i\, p_i(1-p_i)\,]$, so it is
noisier in small tracts; and the predictors are five-year American Community
Survey (ACS) estimates whose margins of error scale as $n_i^{-1/2}$, injecting a
size-dependent errors-in-variables component. Both scale with tract size, so
$\log n_i$ proxies their combined effect. A Koenker--Breusch--Pagan test finds
no dependence of residual variance on the fitted value ($p=0.85$) but strong
dependence on $\log n_i$ ($\mathrm{BP}=145$, $p<10^{-16}$).
Figure~\ref{fig:hetero} shows the relationship directly: residual variance,
computed within tract-size quintiles, falls roughly twofold from the smallest to
the largest tracts and tracks the delta-method prediction. The SDEM curve lies
below the OLS curve---the spatial model reduces the overall variance level---but
runs parallel to it, retaining the same size gradient ($\mathrm{BP}=114$ on the
SDEM residuals): the spatial error model addresses correlation, not
heteroskedasticity.

\FloatBarrier

\begin{figure}[!htbp]
\centering
\includegraphics[width=0.92\textwidth]{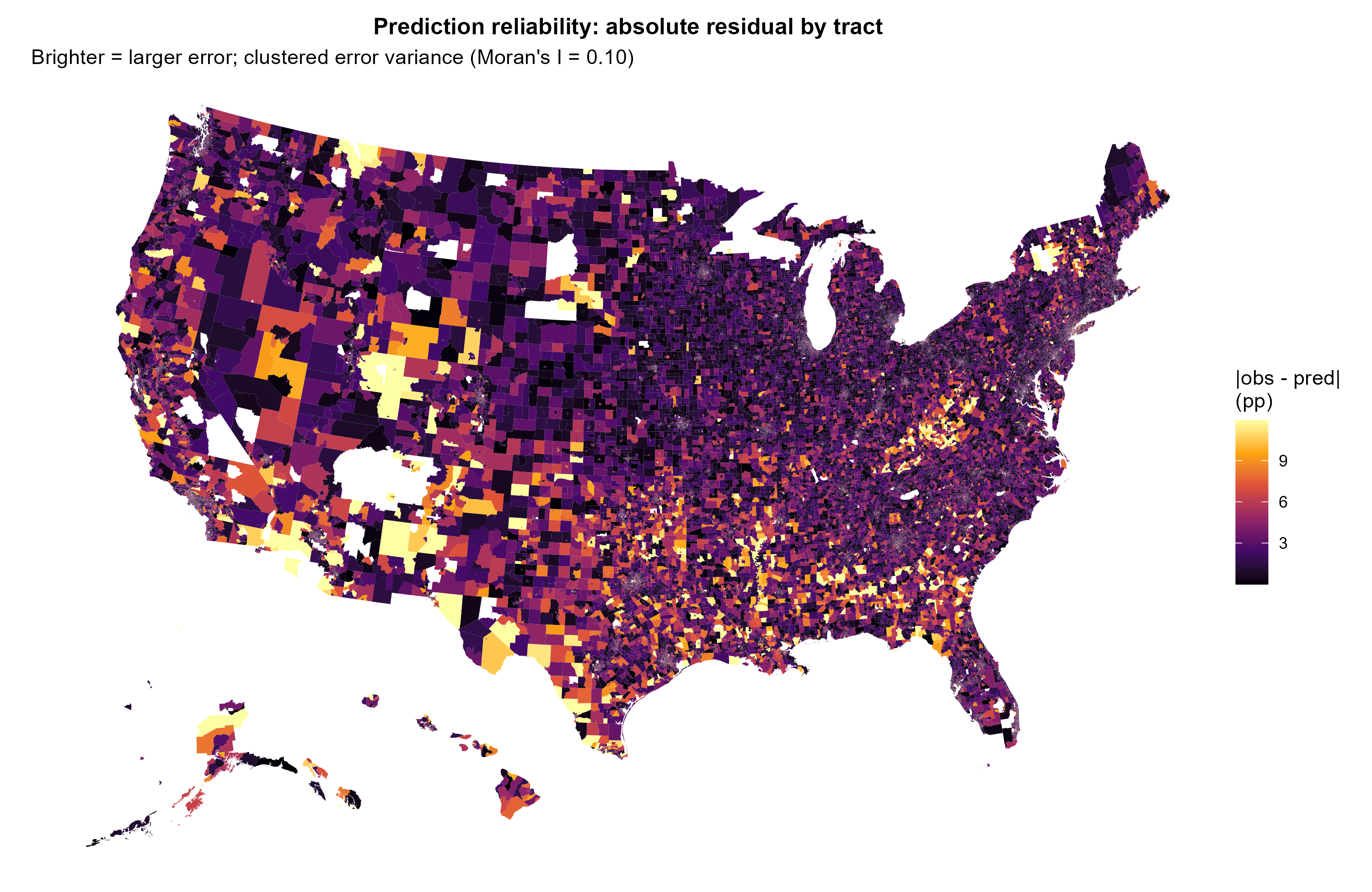}
\caption{Prediction reliability: the absolute SDEM residual (\rateunit) by
tract; brighter tracts have larger errors. Error concentrates in the rural
South and border, the sparsely populated interior West (large-area, small-
population tracts), and scattered urban pockets, while the densely settled East
and Midwest are accurate. The clustered error variance (Moran's $I = 0.10$) is
the spatial footprint of the size-driven heteroskedasticity in
Figure~\ref{fig:hetero}.}
\label{fig:reliability}
\end{figure}

This heteroskedasticity has a spatial footprint. Figure~\ref{fig:reliability}
maps the absolute residual---the size of the prediction error---tract by tract.
Brighter tracts are where the model is least accurate. The errors are not
uniform: they concentrate in the rural South and along the southern border, in
the sparsely populated interior West (where tracts are large in area but small
in population, and therefore noisy), and in scattered urban pockets, while much
of the densely settled East and Midwest is dark (accurate). The squared
residuals are themselves spatially clustered (Moran's $I = 0.096$, $z=45$). The
operational reading is direct and important for a Census instrument: the score
is least reliable precisely in some of the hardest-to-count places, and that
unreliability is a property of where and how small the tracts are, not of the
fitted level.

\FloatBarrier

We bracket the consequences for estimation and inference separately.
Inverse-variance weighted maximum likelihood (weights
$w_i = n_i\, p_i(1-p_i)$) leaves the estimates essentially unchanged:
$\lambda$ shifts by about $0.017$ (SDEM $0.626 \to 0.644$) and the coefficients
move by less than $0.01$ (at most about $0.007$), with no change in sign or
significance and only a negligible change in magnitude. The point estimates and impacts are therefore robust to the
heteroskedasticity. Inference, however, requires correction:
heteroskedasticity-consistent generalized-moments standard errors
\citep{kelejian2010specification} are approximately $1.4$ times the
maximum-likelihood standard errors (SEM $1.36$--$1.50$; SDEM $1.26$--$1.43$),
widening confidence intervals by about forty percent. Given the magnitude of the
effects, this correction overturns no sign conclusion and leaves every
substantively important effect significant; a small number of marginal
coefficients whose in-sample $z$-statistics lie between roughly $2$ and $2.7$
(for example, the neighbor effects of the renter and married-couple shares)
fall below conventional significance once the standard errors are inflated. The exact
per-tract correction, replacing the size proxy with the full ACS predictor
covariance $\Sigma_i$ recovered from the eighty Variance Replicate Tables, is
the subject of a companion measurement-error analysis and is beyond the scope of
this paper.

\subsection{Model selection}
\label{sec:selection}

The evidence supports a three-model account rather than a single winner. The
\textbf{SDEM} is the preferred interpretive model: it absorbs the spatial
structure, its local spillovers admit a direct neighbor-demographic reading
(Figures~\ref{fig:impacts}--\ref{fig:dcneigh}), its coefficient inference is
numerically stable where the lag models are not, and weighting and robust
standard errors leave its conclusions intact. The \textbf{SEM} is its
parsimonious predictive equivalent, marginally best on spatial-block skill and
appropriate when interpretable spillovers are not required. The \textbf{SDM},
although it minimizes AIC, is an in-sample artifact: it is the weakest of the
spatial models out of sample (Figure~\ref{fig:cv}) and rests on a global
feedback mechanism that is difficult to justify substantively for mail
non-response. SARAR is retained only as a diagnostic. We therefore report SDEM
coefficients with heteroskedasticity-robust standard errors as the primary
specification, with SEM and the weighted SDEM as sensitivities; all three agree
on every sign conclusion and on every substantively important effect, with only
a few marginal spillover coefficients changing significance under the robust
standard errors.

\section{Discussion}
\label{sec:discussion}

The contribution of this paper is applied rather than purely methodological.
The spatial econometric tools used here are standard; the novelty is their
application to the Bureau's interpretable LRS framework, the mechanism-specific
interpretation of the residual dependence, and the demonstration that
spatial-block validation changes the model-selection conclusion. In sample, the
SDM appears strongest. Under the validation regime most relevant to transfer
across geography, the error-family models are stronger. That reversal is the
central empirical lesson: for spatial response models, the model that best
explains the map already observed need not be the model that best predicts a
region withheld from estimation.

Substantively, the results argue against a strong global response-contagion
interpretation. The robust diagnostics, SARAR pathology, and spatial-block
validation all point toward error-type dependence and local neighborhood
context. This distinction matters for official statistics. A global lag model
suggests a self-reinforcing process in which response behavior propagates
through the tract network. The evidence here is more consistent with shared
unmeasured context and immediate-neighbor covariate information. The SDEM is
therefore useful not because it produces the lowest possible AIC, but because it
offers a disciplined way to separate own-tract effects from local neighborhood
effects while avoiding the stronger global-feedback interpretation of the SDM.

The operational gains are more nuanced than a simple ``spatial models win''
statement. Under spatial-block validation, SEM and SDEM improve RMSE over OLS
by only about a tenth of a percentage point. Under random folds, where local
neighbors remain available, the spatial gains are much larger. This pattern
does not make the spatial models unimportant; it clarifies their use case. They
are strongest for interpolation within an already observed geographic field and
less transformative for transfer to a wholly withheld region. For a planning
instrument, reporting both regimes is more informative than reporting either
alone.

\section{Limitations and Future Work}
\label{sec:limitations}

Three limitations should guide interpretation. First, the analysis uses the
2010 mail non-return outcome because it is the target of the original LRS and
therefore the cleanest setting for testing what spatial structure remains after
the official twenty-five-predictor OLS specification. The 2020 self-response
environment differs in important ways, including internet response and revised
contact strategies. The next empirical step is a direct replication on the
2020-based Planning Database LRS once the necessary outcome, predictor, and
geographic crosswalk choices are fixed.

Second, this paper treats the ACS predictor estimates as fixed regressors except
for the tract-size heteroskedasticity analysis. That is adequate for the model
comparison question, but a complete uncertainty analysis should propagate ACS
sampling error through the spatial model using the replicate variance
structure. The companion measurement-error analysis noted above is the natural
place for that extension.

Third, the cross-validation design withholds geography, not time. Spatial-block
folds are a demanding test of geographic transfer, but they do not answer how a
model trained before a census performs under a new response regime. Future work
should combine spatial validation with temporal validation across census
vintages, especially if the same modeling logic is applied to the 2020 LRS or to
future self-response scores.

\section{Conclusion}
\label{sec:conclusion}

The original LRS framework remains valuable because it is transparent and
operationally usable. Its residuals, however, contain substantial spatial
structure. Modeling that structure shows that the dependence is primarily
error-type, that local neighbor covariates carry interpretable information, and
that in-sample fit can select a spatial mechanism that does not generalize best
to withheld geography. For peer-reviewed evaluation of response-propensity
scores, the main recommendation is straightforward: spatial models should be
judged with spatial validation, and the substantive interpretation of the
spatial process should be treated as part of model selection rather than an
afterthought.


\section*{Supplementary Material:\\
	A Practical Guide to the Spatial Regression Models}
	
	This supplement is a self-contained tutorial on every method used in the main
	paper. It assumes familiarity with ordinary least squares (OLS) but not with
	spatial econometrics. Each section gives the intuition first, then the model,
	then how to read the output, then the exact \texttt{R} call (using the
	\texttt{spdep} and \texttt{spatialreg} packages). Cross-references to the main
	paper indicate where each method is applied.

	\section{The problem: why ordinary regression is not enough}
	\label{sup:problem}
	
	OLS assumes the residuals are independent: knowing the error in one tract tells
	you nothing about the error in the next. For data tied to a map, that assumption
	usually fails. Neighboring census tracts share a labor market, a media market, a
	local government, a culture of civic participation, and countless unmeasured
	features that no covariate fully captures. As a result, two adjacent tracts tend
	to have similar outcomes \emph{beyond} what their measured characteristics
	explain, and their regression residuals are correlated.
	
	This correlation is called \emph{spatial autocorrelation}. Its consequences
	depend on its source. If only the errors are spatially correlated, OLS stays
	unbiased but becomes inefficient and---more dangerously---its standard errors
	are too small, so coefficients look more significant than they are. If instead
	the outcome itself depends on neighboring outcomes, OLS is also \emph{biased},
	because it omits a relevant (and endogenous) variable. Either way, ignoring the
	dependence is costly. The methods in this guide either remove the
	autocorrelation, exploit it for better prediction, or both. The unifying
	question they answer is: \emph{where does the spatial dependence come from, and
		what should we do about it?}
	
	\section{Spatial weights: defining ``neighbors''}
	\label{sup:weights}
	
	Every spatial model needs a rule for who is a neighbor of whom. That rule is
	encoded in a \emph{spatial weights matrix} $\W$, an $n\times n$ matrix whose
	entry $w_{ij}$ is positive if tracts $i$ and $j$ are neighbors and zero
	otherwise (a tract is never its own neighbor, so $w_{ii}=0$).
	
	Two common rules:
	\begin{itemize}
		\item \textbf{Contiguity (queen).} $w_{ij}>0$ if tracts $i$ and $j$ share any
		border or corner---the moves a queen makes in chess. This is the natural choice
		for irregular polygons like tracts and is the paper's primary specification.
		\item \textbf{$k$-nearest neighbors ($k$NN).} Each tract's neighbors are the $k$
		closest tract centroids. Every tract then has exactly $k$ neighbors, which is
		useful as a robustness check because it removes the variation in the number of
		neighbors that contiguity produces.
	\end{itemize}
	
	\paragraph{Row-standardization.} It is standard to scale each row of $\W$ to sum
	to one, so that $w_{ij}=1/(\text{number of }i\text{'s neighbors})$. With this
	``style \texttt{W}'' scaling, the key quantity $\W y$---the matrix product of
	$\W$ and the outcome vector---is simply the \emph{average of each tract's
		neighbors' outcomes}. This ``spatially lagged'' variable is the building block of
	every model below: $\W y$ is the neighbor-average outcome, and $\W\X$ is the
	neighbor-average of each predictor.
	
	\begin{rcode}
		library(spdep)
		nb <- poly2nb(tracts, queen = TRUE)   # contiguity neighbour list
		W  <- nb2listw(nb, style = "W")       # row-standardised weights
		
		# k-nearest-neighbour alternative (robustness):
		knn <- knn2nb(knearneigh(coords, k = 8))
		Wk  <- nb2listw(knn, style = "W")
	\end{rcode}
	
	\section{Measuring spatial dependence: Moran's \texorpdfstring{$I$}{I}}
	\label{sup:moran}
	
	Before modeling, we test whether spatial dependence is present at all. Moran's
	$I$ is the workhorse statistic. Intuitively, it asks: \emph{do similar values
		cluster together on the map?}---much as one might ask whether expensive houses
	tend to sit near other expensive houses.
	
	Formally, $I$ is a correlation between each value and the average of its
	neighbors' values,
	\[
	I \;=\; \frac{n}{\sum_{i,j} w_{ij}}\,
	\frac{\sum_{i,j} w_{ij}\,(y_i-\bar y)(y_j-\bar y)}{\sum_i (y_i-\bar y)^2}.
	\]
	It runs roughly from $-1$ to $+1$. Values near $0$ mean no spatial pattern
	(the map could have been shuffled at random); positive values mean clustering
	(high near high, low near low); negative values mean a checkerboard.
	
	The most useful version applies $I$ to the \emph{residuals} of a fitted model.
	A large residual $I$ says the model has left spatial structure on the table; an
	$I$ near zero says the spatial structure has been absorbed. In the paper, OLS
	residuals have $I=0.399$ (strong clustering), while the error and Durbin models
	drive it to about $-0.04$ (Table~1 of the main text).
	
	\begin{rcode}
		moran.test(y, W)            # clustering in a raw variable
		lm.morantest(ols_fit, W)    # clustering left in OLS residuals
	\end{rcode}
	
	\section{A family of spatial models}
	\label{sup:family}
	
	All the models in the paper are special cases of one general equation. Writing
	$y$ for the outcome, $\X$ for the predictors, $u$ for the error, and
	$\varepsilon$ for pure noise:
	\begin{equation}
		y \;=\; \underbrace{\rho\,\W y}_{\text{lagged outcome}}
		\;+\; \X\beta
		\;+\; \underbrace{\W\X\,\theta}_{\text{lagged predictors}}
		\;+\; u,
		\qquad
		u \;=\; \underbrace{\lambda\,\W u}_{\text{lagged error}} \;+\; \varepsilon .
		\label{eq:gns}
	\end{equation}
	The three spatial terms are governed by three parameters. Setting some of them to
	zero produces each named model (Table~\ref{tab:zoo}). Before the catalogue, two
	distinctions make sense of the whole family.
	
	\paragraph{Distinction 1: lag versus error.}
	\begin{itemize}
		\item A \textbf{lag} term ($\rho\,\W y$) says a tract's \emph{outcome} depends on
		its neighbors' outcomes. This is a substantive interaction---``contagion.'' If
		low civic participation genuinely diffuses across a neighborhood, the outcome
		itself spills over.
		\item An \textbf{error} term ($\lambda\,\W u$) says the \emph{unmeasured shocks}
		are spatially correlated. This is a nuisance: a regional outreach campaign, a
		shared media market, or a cultural factor that we never measured affects
		adjacent tracts together. The outcome does not spill over; the omitted variables
		do.
	\end{itemize}
	The distinction is not cosmetic. If the truth is an error process and you fit a
	lag model, you will report spurious spillovers. The paper's diagnostics point
	firmly to an error process.
	
	\paragraph{Distinction 2: local versus global spillovers.}
	\begin{itemize}
		\item \textbf{Lagged predictors} ($\W\X\,\theta$) create \emph{local} spillovers:
		a tract is affected by its immediate neighbors' characteristics, and the effect
		stops there. One ring of ripples.
		\item \textbf{A lagged outcome} ($\rho\,\W y$) creates \emph{global} spillovers.
		Because $y$ appears on both sides, solving for it gives
		$y=(\mathbf{I}-\rho\W)^{-1}(\X\beta+\dots)$, and the multiplier
		$(\mathbf{I}-\rho\W)^{-1}=\mathbf{I}+\rho\W+\rho^2\W^2+\cdots$ propagates any
		change through neighbors, neighbors-of-neighbors, and back again as feedback.
		Ripples across the whole pond.
	\end{itemize}
	
	\begin{table}[t]
		\centering
		\caption{The model family as restrictions of equation~\eqref{eq:gns}.
			A bullet ($\bullet$) marks a term that is present. ``Spillovers'' describes how a
			change in one tract's predictor reaches others.}
		\label{tab:zoo}
		\small
		\begin{tabular}{llcccll}
			\toprule
			Model & Name & $\rho\W y$ & $\W\X\theta$ & $\lambda\W u$ & Spillovers & \texttt{R} function \\
			\midrule
			OLS   & Ordinary least squares      &          &          &          & none           & \texttt{lm} \\
			SLX   & Spatial lag of $X$          &          & $\bullet$&          & local          & \texttt{lmSLX} \\
			SLM   & Spatial lag (SAR)           & $\bullet$&          &          & global         & \texttt{lagsarlm} \\
			SEM   & Spatial error               &          &          & $\bullet$& none (in mean) & \texttt{errorsarlm} \\
			SDM   & Spatial Durbin              & $\bullet$& $\bullet$&          & global         & \texttt{lagsarlm}$^\dagger$ \\
			SDEM  & Spatial Durbin error        &          & $\bullet$& $\bullet$& local          & \texttt{errorsarlm}$^\dagger$ \\
			SARAR & Combined lag+error (SAC)    & $\bullet$&          & $\bullet$& global         & \texttt{sacsarlm} \\
			\bottomrule
		\end{tabular}
		
		\smallskip
		{\footnotesize $^\dagger$With \texttt{Durbin = TRUE} to add the lagged
			predictors $\W\X\theta$.}
	\end{table}
	
	\subsection{OLS --- the baseline}
	$y=\X\beta+\varepsilon$. No spatial terms. Fast and familiar, but if the
	residuals are spatially autocorrelated (Section~\ref{sup:moran}) its standard
	errors are untrustworthy. It is the null against which the spatial models are
	judged.
	
	\subsection{SLX --- spatial lag of \texorpdfstring{$X$}{X}}
	$y=\X\beta+\W\X\theta+\varepsilon$. Add the neighbors' average predictors as
	extra regressors. This is the simplest way to let a tract be influenced by its
	surroundings, and its spillovers are \emph{local} and read directly off
	$\theta$. It is estimated by OLS (the added regressors are just more columns), so
	it needs no special machinery---but on its own it rarely removes residual
	autocorrelation, because it has no autoregressive term.
	
	\begin{rcode}
		library(spatialreg)
		slx <- lmSLX(formula, data = d, listw = W)
	\end{rcode}
	
	\subsection{SLM --- spatial lag (SAR)}
	$y=\rho\,\W y+\X\beta+\varepsilon$. The outcome depends on neighbors' outcomes.
	Use it when you believe in genuine \emph{contagion} of the outcome. Its
	spillovers are global through the multiplier $(\mathbf{I}-\rho\W)^{-1}$, so its
	coefficients are not marginal effects and must be converted to impacts
	(Section~\ref{sup:impacts}). A limitation: the lag model forces every predictor
	to share the same spillover pattern, which is often implausible.
	
	\begin{rcode}
		slm <- lagsarlm(formula, data = d, listw = W)
	\end{rcode}
	
	\subsection{SEM --- spatial error}
	$y=\X\beta+u$, $u=\lambda\,\W u+\varepsilon$. The mean is ordinary; only the
	errors are spatially correlated. Use it when the dependence comes from omitted,
	spatially smooth variables rather than from outcome contagion. SEM does not
	change the meaning of $\beta$ (there are no spillovers in the mean); it makes
	estimation efficient and the standard errors honest. In the paper this is the
	parsimonious model that the diagnostics favor.
	
	\begin{rcode}
		sem <- errorsarlm(formula, data = d, listw = W)
	\end{rcode}
	
	\subsection{SDM --- spatial Durbin}
	$y=\rho\,\W y+\X\beta+\W\X\theta+\varepsilon$. The lag model plus lagged
	predictors. It is the most flexible global-spillover model and often fits best
	in sample, but that flexibility can overfit: in the paper SDM minimizes AIC yet
	predicts worst out of sample (Section~5 of the main text). Its spillovers are
	global and require the impacts calculation.
	
	\begin{rcode}
		sdm <- lagsarlm(formula, data = d, listw = W, Durbin = TRUE)
	\end{rcode}
	
	\subsection{SDEM --- spatial Durbin error}
	$y=\X\beta+\W\X\theta+u$, $u=\lambda\,\W u+\varepsilon$. Lagged predictors for
	\emph{local, interpretable} spillovers, plus a spatial error term to mop up the
	remaining omitted-variable correlation. It combines the two pieces the
	diagnostics call for---neighbor effects in the mean and correlated errors---
	without the hard-to-justify global feedback of a lagged outcome. This is the
	paper's preferred interpretive model. Its direct effects are $\beta$ and its
	indirect (spillover) effects are simply $\theta$.
	
	\begin{rcode}
		# "emixed" and Durbin=TRUE are equivalent ways to request lagged X:
		sdem <- errorsarlm(formula, data = d, listw = W, Durbin = TRUE)
	\end{rcode}
	
	\subsection{SARAR --- combined lag and error (SAC)}
	$y=\rho\,\W y+\X\beta+u$, $u=\lambda\,\W u+\varepsilon$. Both a lagged outcome and
	a lagged error. In principle the most general of the common models, but the two
	spatial parameters often compete for the same signal and become weakly
	identified. The paper sees exactly this: SARAR returns a negative, implausible
	$\rho$ alongside a large $\lambda$, the classic sign of a spurious lag in an
	error-driven process, so it is kept only as a diagnostic.
	
	\begin{rcode}
		sac <- sacsarlm(formula, data = d, listw = W)
	\end{rcode}
	
	\section{Choosing a model: specification tests}
	\label{sup:tests}
	
	Two complementary strategies pick a model: build up from OLS (specific-to-
	general) and trim down from a richer model (general-to-specific).
	
	\subsection{Specific-to-general: Lagrange-multiplier (Rao-score) tests}
	These tests examine the OLS residuals and ask whether adding a lag term or an
	error term would help---\emph{without} having to estimate the larger models. The
	plain LM tests for lag and for error are often both significant, because each is
	contaminated by the other. The \emph{robust} versions correct for that: the
	robust-error test asks whether an error term is needed \emph{given} a possible
	lag, and vice versa. Whichever robust statistic is far larger points to the
	process to adopt. In the paper the robust-error statistic dwarfs the robust-lag
	statistic ($\sim$26-fold under contiguity), nominating an error process.
	
	\begin{rcode}
		# spdep >= 1.3 renamed lm.LMtests() to lm.RStests();
		# look at the *robust* rows (adjRSerr, adjRSlag).
		lm.RStests(ols_fit, W, test = "all")
	\end{rcode}
	
	\subsection{General-to-specific: likelihood-ratio, common-factor, and Hausman tests}
	Having estimated the richer models, nested \textbf{likelihood-ratio (LR)} tests
	check whether each added term earns its place: does the error term improve on
	OLS? do the lagged predictors improve on the error model? In the paper every
	such test rejects decisively.
	
	The \textbf{common-factor (COMFAC) test} \citep{burridge1981testing} asks a
	sharper question: is the spatial Durbin model just a spatial error model in
	disguise? Algebraically, SDM reduces to SEM exactly when $\theta=-\rho\beta$
	(the ``common factor'' restriction). Rejecting it means the lagged predictors
	carry information a pure error model cannot reproduce.
	
	The \textbf{spatial Hausman test} \citep{pace2008spatial} compares the OLS and
	SEM coefficient vectors. If a pure error model were correctly specified the two
	should agree; a large discrepancy signals that something systematic---here, the
	neighbor covariates---is still missing, motivating the Durbin terms.
	
	\begin{rcode}
		LR.Sarlm(sdem, sem)    # do the Durbin terms improve on the pure error model?
		Hausman.test(sem)      # OLS vs SEM coefficient agreement
	\end{rcode}
	
	\section{Interpreting the results: impacts}
	\label{sup:impacts}
	
	In any model with $\W y$ or $\W\X$, a coefficient is no longer the marginal
	effect of a predictor. Changing predictor $k$ in one tract changes that tract's
	own outcome (a \textbf{direct} effect) and, through the spatial terms, other
	tracts' outcomes (an \textbf{indirect} or spillover effect); their sum is the
	\textbf{total} effect \citep{lesage2009introduction}.
	
	How the indirect effect is computed depends on the model, and this is where the
	local-versus-global distinction bites:
	\begin{itemize}
		\item For \textbf{SLX and SDEM} (local spillovers), the indirect effect of
		predictor $k$ is just its lagged coefficient $\theta_k$. Simple and transparent.
		\item For \textbf{SLM and SDM} (global spillovers), the effects run through the
		multiplier $(\mathbf{I}-\rho\W)^{-1}$ and accumulate across the map, so they are
		larger and are summarized by averaging the rows of the resulting effect matrix.
		Confidence intervals are obtained by simulation.
	\end{itemize}
	This is why, in the paper, the SDM and SDEM \emph{direct} effects coincide while
	the SDM \emph{indirect} effects are several times larger: the difference is the
	global multiplier, not extra structure in the data. Reporting impacts---not raw
	coefficients---is essential for any Durbin or lag model.
	
	\begin{rcode}
		# direct / indirect / total with simulated z-statistics and intervals:
		summary(impacts(sdem, listw = W, R = 200), zstats = TRUE)
		summary(impacts(sdm,  listw = W, R = 200), zstats = TRUE)
	\end{rcode}
	
	\section{Honest standard errors under heteroskedasticity}
	\label{sup:robust}
	
	Maximum-likelihood estimation of these models assumes the innovation
	$\varepsilon$ has constant variance. When it does not---as here, where the
	logit-transformed rate is noisier in small tracts and the predictors carry
	size-dependent survey error---the point estimates remain reliable but the
	model-based standard errors are too small.
	
	The fix is a heteroskedasticity-consistent estimator of the standard errors
	\citep{kelejian2010specification}, computed by generalized method of moments
	(GM) and valid without assuming constant variance. In the paper these robust
	standard errors are about $1.4$ times the maximum-likelihood values; the
	estimates and the substantively important conclusions are unchanged, though a
	few marginal spillovers lose significance once the intervals are widened
	appropriately.
	
	\begin{rcode}
		library(sphet)
		spreg(formula, data = d, listw = W, model = "error", het = TRUE)
	\end{rcode}
	
	\section{Prediction and validation}
	\label{sup:prediction}
	
	\paragraph{Spatial prediction.} Predicting a held-out tract is not just
	$\X\hat\beta$. A fitted spatial model decomposes the outcome into a
	\emph{trend} (the covariate part $\X\hat\beta$, plus $\W\X\hat\theta$ for Durbin
	models) and a \emph{signal} (the spatial part learned from neighbors). The best
	prediction adds the signal that the tract's \emph{observed} neighbors imply, via
	the optimal predictors of \citet{goulard2017predictions}. A tract surrounded by
	high-non-return neighbors is predicted higher than its covariates alone would
	suggest---which is exactly the value a spatial model adds.
	
	\begin{rcode}
		# predict() implements the Goulard et al. (2017) predictors via pred.type;
		# the trend-and-signal types borrow the held-out tract's neighbours
		# (see ?predict.Sarlm for the full set, e.g. "TC", "BP"):
		predict(sem, newdata = test_tracts, listw = W_full, pred.type = "TC")
	\end{rcode}
	
	\paragraph{Why spatial cross-validation.} Ordinary $k$-fold cross-validation
	scatters the held-out tracts among the training tracts. Because of spatial
	autocorrelation, almost every test tract then sits next to a training tract, and
	the model can ``peek'' at the answer through its neighbors---inflating apparent
	skill. \emph{Spatial-block} cross-validation instead holds out whole contiguous
	regions, so a test tract's neighbors are withheld with it. The gap between the
	two regimes measures how much a model leans on local neighbors (which it cannot
	do for genuinely new geography) versus transferable covariate structure. In the
	paper this gap is what exposes the in-sample-best SDM as a comparatively poor
	out-of-sample predictor.
	
	\begin{rcode}
		# build contiguous folds by clustering tract centroids:
		set.seed(42)
		folds <- kmeans(scale(coords), centers = 5, nstart = 10)$cluster
		# then hold out each fold's region in turn, refit on the rest, and predict.
	\end{rcode}
	
	\section{A short decision guide}
	\label{sup:guide}
	
	\begin{enumerate}
		\item \textbf{Test first.} Compute residual Moran's $I$ on the OLS fit. If it is
		near zero, ordinary regression is adequate.
		\item \textbf{Lag or error?} Read the robust LM/RS tests. A dominant robust-error
		statistic points to an error model; a dominant robust-lag statistic points to a
		lag model.
		\item \textbf{Local spillovers wanted?} If the neighbors' \emph{characteristics}
		should matter and you want interpretable, local spillover estimates, add lagged
		predictors---SLX (if errors are clean) or SDEM (if they are not).
		\item \textbf{Global contagion of the outcome?} Only if there is a substantive
		reason for the \emph{outcome} to spill over should you adopt a lagged-outcome
		model (SLM/SDM), and then report impacts, not coefficients.
		\item \textbf{Check robustness.} Re-estimate under an alternative weights matrix,
		use heteroskedasticity-robust standard errors, and---above all---validate out of
		sample with spatial-block folds, since in-sample fit can mislead.
	\end{enumerate}
	
	For the Low Response Score, this path leads to an error-type process with
	interpretable local spillovers and correlated errors: the spatial Durbin error
	model, reported with robust standard errors and validated on held-out regions.


\begin{thebibliography}{12}
\bibitem[Anselin et al.(1996)]{anselin1996simple}
Anselin, L., Bera, A.~K., Florax, R., and Yoon, M.~J. (1996).
Simple diagnostic tests for spatial dependence.
\textit{Regional Science and Urban Economics}, 26(1), 77--104.

\bibitem[Burridge(1981)]{burridge1981testing}
Burridge, P. (1981).
Testing for a common factor in a spatial autoregression model.
\textit{Environment and Planning A}, 13(7), 795--800.

\bibitem[U.S. Census Bureau(2026)]{census2026pdbupdate}
U.S. Census Bureau. (2026).
Updating the Planning Database with the latest socioeconomic, demographic and
housing data.
\textit{Random Samplings Blog}.
\url{https://www.census.gov/newsroom/blogs/random-samplings/2026/03/updating-the-planning-database.html}.

\bibitem[Elhorst(2010)]{elhorst2010applied}
Elhorst, J.~P. (2010).
Applied spatial econometrics: raising the bar.
\textit{Spatial Economic Analysis}, 5(1), 9--28.

\bibitem[Erdman and Bates(2017)]{erdman2017lrs}
Erdman, C., and Bates, N. (2017).
The Low Response Score (LRS): A metric to locate, predict, and manage
hard-to-survey populations.
\textit{Public Opinion Quarterly}, 81(1), 144--156.

\bibitem[Goulard et al.(2017)]{goulard2017predictions}
Goulard, M., Laurent, T., and Thomas-Agnan, C. (2017).
About predictions in spatial autoregressive models: optimal and almost optimal
strategies.
\textit{Spatial Economic Analysis}, 12(2--3), 304--325.

\bibitem[Kelejian and Prucha(2010)]{kelejian2010specification}
Kelejian, H.~H., and Prucha, I.~R. (2010).
Specification and estimation of spatial autoregressive models with
autoregressive and heteroskedastic disturbances.
\textit{Journal of Econometrics}, 157(1), 53--67.

\bibitem[LeSage and Pace(2009)]{lesage2009introduction}
LeSage, J., and Pace, R.~K. (2009).
\textit{Introduction to Spatial Econometrics}.
Chapman and Hall/CRC, Boca Raton.

\bibitem[Pace and LeSage(2008)]{pace2008spatial}
Pace, R.~K., and LeSage, J.~P. (2008).
A spatial Hausman test.
\textit{Economics Letters}, 101(3), 282--284.

\bibitem[Roberts et al.(2017)]{roberts2017cross}
Roberts, D.~R., et al. (2017).
Cross-validation strategies for data with temporal, spatial, hierarchical, or
phylogenetic structure.
\textit{Ecography}, 40(8), 913--929.


\bibitem[Burridge(1981)]{burridge1981testing}
Burridge, P. (1981). Testing for a common factor in a spatial autoregression
model. \textit{Environment and Planning A}, 13(7), 795--800.




\bibitem[Pace and LeSage(2008)]{pace2008spatial}
Pace, R.~K., and LeSage, J.~P. (2008). A spatial Hausman test.
\textit{Economics Letters}, 101(3), 282--284.
\end{thebibliography}
\end{document}